\documentstyle[12pt]{article}
\def\laq{\raise 0.4 ex \hbox{$<$}\kern -0.8 em\lower 0.62 ex\hbox{$\sim$}}
\def\gaq{\raise 0.4 ex \hbox{$>$}\kern -0.7 em\lower 0.62 ex\hbox{$\sim$}}
 
\def\beq{\begin{equation}}
\def\eeq{\end{equation}}
\def\bea{\begin{eqnarray}}
\def\eea{\end{eqnarray}}
\def\bq{\begin{quote}}
\def\eq{\end{quote}}

\def\ka{\kappa}
\def\la{\lambda}

\def\si{\sigma}

\def\La{\Lambda}

\def\mn{{\mu\nu}}
\def\cl{{\cal L}}

\def\frac#1#2{{\textstyle{{#1}\over {#2}}}}

\def\lsim{\mathrel{\rlap{\lower4pt\hbox{\hskip1pt$\sim$}}
    \raise1pt\hbox{$<$}}}
\def\gsim{\mathrel{\rlap{\lower4pt\hbox{\hskip1pt$\sim$}}
    \raise1pt\hbox{$>$}}}
\def\sqr#1#2{{\vcenter{\vbox{\hrule height.#2pt
         \hbox{\vrule width.#2pt height#1pt \kern#1pt
         \vrule width.#2pt}
         \hrule height.#2pt}}}}
\def\square{\mathchoice\sqr66\sqr66\sqr{2.1}3\sqr{1.5}3}

\def\AJ{{\it Ap. J.} }

\def\CQG{{\it Class. Quantum Gravity} }

\def\GRG{{\it Gen. Relativity and Gravitation} }

\def\IJMP{{\it Int. J. Mod. Phys.} }

\def\MNRAS{{\it Mon. Not. R. Ast. Soc.} }

\def\PL{{\it Phys. Lett.} }
\def\PR{{\it Phys. Rev.} }
\def\PRL{{\it Phys. Rev. Lett.} }

\def\RMP{{\it Rev. Mod. Phys.} }

\def\gappeq{\mathrel{\rlap {\raise.5ex\hbox{$>$}}
{\lower.5ex\hbox{$\sim$}}}}
 
\def\lappeq{\mathrel{\rlap{\raise.5ex\hbox{$<$}}
{\lower.5ex\hbox{$\sim$}}}}

\begin{document}

\begin{center}
{{\bf Using a non-minimal coupling between matter and curvature to sequester the Cosmological Constant}}

\vglue 0.5cm
{Orfeu Bertolami$^{1,2}$, J. P\'aramos$^{1}$}

\bigskip
{$^{1}$ Departamento de F\'\i sica e Astronomia, Faculdade de Ci\^encias, Universidade do Porto\\}
\smallskip
{Rua do Campo Alegre 687, 4169-007 Porto, Portugal\\}

\vglue 0.3cm

{$^{2}$ Centro de F\'\i sica do Porto, Faculdade de Ci\^encias, Universidade do Porto \\}
\smallskip
{Rua do Campo Alegre 687, 4169-007 Porto, Portugal\\}
\end{center}

\vglue 0.7cm

\setlength{\baselineskip}{0.7cm}

\centerline{\bf  Abstract}
\vglue 0.7cm
\noindent

We present a novel mechanism for generating a Cosmological Constant and suitably sequestering the vacuum contribution to it, so that the eponymous Cosmological Constant problem is avoided.

We do so by resorting to a model endowed with a non-minimal coupling between curvature and matter in an appropriately defined relaxed regime, and show that this shares features with both Unimodular gravity as well as a recent proposal to sequester the vacuum contribution through the use of an external term to the action functional.


\section{Introduction}

About twenty years have passed since one of the most striking discoveries of modern Cosmology, namely that matter does not dominate the dynamics of the Universe, slowly braking its expansion and perhaps leading to an ensuing collapse, but that the expansion is actually accelerating, thus requiring a source in the Einstein field equations endowed with negative pressure \cite{accexp}.

This acceleration can be attained by resorting to a Cosmological Constant (CC), first posited by Einstein one century ago to avoid the expansion of the Universe, by considering a quintessencial scalar field slow-rolling down a suitable potential \cite{quintessence} or an exotic equation of state such as the (generalised) Chaplygin gas \cite{Chaplygin}.

However, these lead to the eponymous problem of how to cancel out the difference of $\sim 120$ orders of magnitude between the contribution $\La_0 $ stemming from summing the zero-point energies of Standard Model fields and beyond, and the observed value $\La \sim H_0^2$ (where $H_0 \simeq 10^{-42}$ GeV is the current value of the Hubble parameter) \cite{problem}. In General Relativity (GR), this requires that the `bare' CC is extremely fine-tuned so that it almost exactly cancels out the large value of $\Lambda_0$, leaving as a residue the observed value $\Lambda$.

Even though it does not solve the problem, Unimodular gravity \cite{unimodular} distinguishes itself amongst several proposals for generating a CC in an elegant way, as it arises out of an {\it a priori} constraint on the allowed diffeomorphisms. In another proposal relevant for our discussion \cite{sequester}, the required sequester of the ``bare'' and vacuum-energy contributions to the CC is achieved consideering a modification of GR which implies the addition of global terms to the Einstein-Hilbert action, together with an adequate scalar field acting as a Lagrange multiplier. 

This work aims to show that such a sequester may also be obtained by resorting to a non-minimal coupling between matter and curvature \cite{NMC}, through a mechanism that shares significant features with Unimodular gravity. Furthermore, such a model shares some common features with the so-called emergent gravity models \cite{emergent1} (see {\it e.g.} Ref. \cite{emergent2} for an example of an observational signature), but arises out of the usual action functional formalism.

\section{Relaxed non-minimal coupling between curvature and matter}

We now consider a model with a generalised non-minimal coupling (NMC) between curvature and matter \cite{NMC}, of the form
\begin{equation} \label{NMCaction} S = \int d^4x \sqrt{-g} \left[ \ka f_1(R) + f_2(R) \cl(g^\mn,\chi) \right]~~. \end{equation}
where $\kappa=c^4/(16\pi G)$, $f_1(R)$ and $f_2(R)$ are generic functions of the scalar curvature, $\chi$ denotes matter fields and $\cal{L}$ their Lagrangian density. This encompasses so-called $f(R)$ theories, one of the outstanding proposals of the so-called `dark gravity' type \cite{fR}; the additional NMC leads to further phenomenological implications such as the mimicking of cluster and galactic dark matter \cite{NMCdm}, dark energy \cite{NMCcosmo1} and the CC \cite{NMCCC} (see Ref. \cite{reviewNMC} for a thorough review). It should be noted that by enhancing the gravitational field created by matter, this model can account for the Tully-Fisher law \cite{NMC,NMCdm}.

Variation with respect to the metric leads to the field equations:
\begin{equation} \label{NMCfieldeqs} \left(f'_1 + {1 \over \ka} f'_2 \cl\right) R_\mn -  {1 \over 2}g_\mn f_1 = {1 \over 2 \ka} f_2 T_\mn + (\nabla_\mu \nabla_\nu - g_\mn \square) \left(f'_1 + {1 \over \ka} f'_2 \cl \right) ~~.\end{equation}
\noindent With the contracted Bianchi identities one can show that the energy-momentum tensor of matter is no longer covariantly conserved:
\begin{equation} \label{NMCnoncons} \nabla_\mu T^\mn = {f'_2 \over f_2} \left( g^\mn \cl - T^\mn \right) \nabla_\mu R~~, \end{equation}
\noindent a central feature of the model \cite{NMC,noncons,noncons2}.

By performing a conformal transformation $g_\mn \to \psi g_\mn$, the model above translates into the equivalent action,
\begin{equation} \label{NMCequivaction} S = \int d^4x \sqrt{-g} \left[ \ka R + { \ka \over \psi^2} \left( f_1(\phi) - \psi \phi - {3 \over 2} \square\psi \right) + f_2(\phi) \cl\left({g^\mn \over \psi},\chi\right) \right] ~~,\end{equation}
with a physical metric $g^\mn / \psi$ coupling to matter fields $\chi$ \cite{scalarNMC}. Variation of the action with respect to the two scalar fields yields the dynamical identification 
\begin{equation} \psi = f'_1(R) + {1 \over \ka} f'_2(R) \cl ~~~~,~~~~\phi = R~~.\end{equation}
While $\phi$ acts as an auxiliary field with no kinetic term, the scalar field $\psi$ embodies an additional scalar degree of freedom, as found in $f(R)$ theories \cite{scalarfR}.

As firstly explored in Refs. \cite{NMCcosmo1}, the so-called relaxed regime 
\begin{equation} \label{condition} \psi = 1 \to \ka f'_1(R) + f'_2(R) \cl = \ka ~~,\end{equation}
naturally arises out of a dynamical system formulation of the above system of differential equations \cite{NMCcosmo2}, and can be interpreted as an asymptotic regime for this dynamical scalar field, instead of being imposed as an {\it a priori} constraint.

Although the {\it r.h.s.} can be set to any constant value through a suitable conformal transformation, we adopt the choice $\psi = 1$ as it is satisfied by GR, where $f'_1(R) = 1 $ and $f'_2(R) = 0$. In fact, the above constraint establishes a class of models, and GR may be considered the simplest of these: for more convoluted choices of the functions $f_1(R)$ and $f_2(R)$, the condition Eq. (\ref{condition}) is attained only asymptotically, {\it i.e.} it acts as an attractive fixed point for the cosmological equations, as shown in Ref. \cite{NMCcosmo2}.

By the same token, the mechanism outlined below should not be considered to extend to astrophysical or local scales, as in these scenarios the required fixed point condition may not have been attained --- although a similar model indeed resorts to condition Eq. (\ref{condition}) in order to account for galactic dark matter \cite{NMCdm}.

Inserting condition (\ref{condition}) into the trace of Eq. (\ref{NMCfieldeqs}) leads to
\begin{equation}
\label{tracecondition} 2f_1 = R - { 1 \over 2\ka } f_2 T ,
\end{equation}
so that Eq. (\ref{NMCfieldeqs}) read
\begin{equation} \label{NMCunimodular} R_\mn - {1 \over 4} g_\mn R = {f_2(R) \over 2\ka} \left[ T_\mn - {1 \over 4} g_\mn T\right] ~~, \end{equation}
\noindent where $T$ is the trace of the energy-momentum tensor $T_\mn$.

In the case of a weak NMC $f_2(R) \approx 1$, the above is strikingly similar to the traceless equation of motion of Unimodular Gravity \cite{unimodular}, which stems from varying the Einstein-Hilbert action imposing the condition on the determinant of the metric $\sqrt{-g} = 1$; notice that one cannot simply impose a weak coupling $f_2(R) = 1$, as the dynamical identification with a two-scalar field model would break down and the ensuing cosmological dynamical system would no longer yield the relaxed regime Eq. (\ref{condition}) \cite{scalarfR}.

In Unimodular Gravity, the Bianchi identities $ \nabla^\nu R = 2\nabla_\mu R^\mn $ together with the assumption that the energy-momentum tensor of matter is covariantly conserved, $\nabla_\mu T^\mn = 0$, implies that $( 2\ka R + T)_{,\nu} = 0$, so that $ R + T/2\ka = 4\La_1 $, with $\La_1 $ an integration constant. Substituting back into the Unimodular equation of motion ({\it i.e.} Eq. (\ref{NMCunimodular}) with $f_2(R) = 1$) leads to
\begin{equation} \label{fieldeqs_unimodular2} R_\mn - {1 \over 2} g_\mn R + g_\mn \La = {1 \over 2\ka} T_\mn ~~, \end{equation}
so that a contribution to the CC arises as an integration constant out of the restriction $\sqrt{-g}=1$ on the the allowed diffeomorphisms. However, this result only accounts for a natural generation of $\La_1$, but does not solve the problem of the CC: indeed, one must still fine-tune the latter so that the resulting value $\La = \La_0 + \La_1$ coincides with observations. 

The traceless form of Eq. (\ref{NMCunimodular}) is also similar to the equations of motion derived from a recent attempt to tackle the Cosmological Constant problem \cite{sequester}: this is achieved by supplementing the Einstein-Hilbert action with an external term $\si$, together with an auxiliary (i.e. non-dynamical) scalar field $\la$ non-minimally coupled to matter \cite{sequester},
\begin{equation} \label{action_sequester} S = \int d^4 x \sqrt{-g} \left[ \ka R - \La_1 + \la^4 \cl\left( \la^{-2}g^\mn, \chi\right) \right] + \si \left( {\La_1 \over \la^4 \mu^4} \right)~~, \end{equation}
where $\mu$ is a phenomenological parameter and $\chi$ are matter fields which couple to the `physical' metric $\la^{-2}g^\mn$: the additional coupling of the matter Lagrangian density with the scalar field is of the form $\la^4$ to ensure that the ensuing mechanism is valid even if radiative corrections to the vacuum energy are considered.

Varying the action Eq. (\ref{action_sequester}) with respect to $\La_1$, $\la$ and the metric leads to the field equations:
\begin{equation} \label{fieldeqs_sequester} R_\mn - {1 \over 2} g_\mn R = {1 \over 2\ka} \left( T_\mn - {1 \over 4} g_\mn \left<T\right> \right) ~~, \end{equation}
where $\left<T\right> \equiv \int d^4 x \sqrt{-g} T / \int d^4 x \sqrt{-g} $ is the ``cosmic'' average of $T$.This means that this proposal is non-local (non-locality is also at the heart of some radical proposals to tackle the problem of the CC \cite{CCproposals}).

Following the decomposition Eq. (\ref{fieldeqs_sequester}) of the Lagrangian density into vacuum energy plus matter contributions, we see that the vacuum contribution to the above vanishes, since the  vacuum-energy $\La_0$ and the model parameter $\La_1$ equal their cosmic average, $\left<\La_i \right> = \La_i $. Thus, Eq. (\ref{fieldeqs_sequester}) reads
\begin{equation} \label{fieldeqs_sequester2} R_\mn - {1 \over 2} g_\mn R + g_\mn \La = {1 \over 2\ka} \tau_\mn ~~, \end{equation}
so that the constant term appearing, identified with the observed value of the CC, $\La = \left<\tau\right>/( 8\ka)$, reflects the cosmic average of regular matter types. Thus, a sequester of the vacuum energy contribution occurs, at the expense of locality.

\section{Results}

Inspired by Unimodular gravity, we apply the covariant derivative to Eq. (\ref{NMCunimodular}) and use the non-conservation law (\ref{NMCnoncons}), in an attempt to derive a conserved quantity that may act as a ``bare'' CC in the field equations:
\begin{eqnarray}
&& \nabla^\nu R = 4\nabla_\mu \left( R^\mn - {1 \over 4} g^\mn R \right) = \\ \nonumber && {2 \over \ka } \left[ \left( T^\mn - {1 \over 4} g^\mn T\right) \nabla_\mu f_2 + f_2 \left(  \nabla_\mu T^\mn - {1 \over 4} g^\mn  \nabla_\mu T\right) \right] = \\ \nonumber && {2 \over \ka } \left[ \left( T^\mn - {1 \over 4} g^\mn T\right) f_2' \nabla_\mu R + f'_2 \left( g^\mn \cl - T^\mn \right) \nabla_\mu R - {1 \over 4} f_2 \nabla^\nu T \right] = \\ \nonumber && {2 \over \ka } \left[ f'_2 \cl \nabla^\nu R - {1 \over 4} \nabla^\nu (f_2 T) \right] =2 \nabla^\nu (R - f_1) - {1 \over 2\ka} \nabla^\nu (f_2 T) \to \\ && \nabla^\nu \left( 2f_1 - R + {1 \over 2\ka } f_2 T \right) = 0, 
\end{eqnarray}
which, given the trace Eq. (\ref{tracecondition}), vanishes trivially.

Thus, in the present scheme no integration constant is obtained from the Bianchi identities; instead, Eq. (\ref{condition}) directly provides the required conserved quantity, with Eq. (\ref{tracecondition}) acting as an additional constraint on the forms $f_1(R)$ and $f_2(R)$.

Since the relaxed regime posited by Eq. (\ref{condition}) is a fixed point of the cosmological dynamical system derived from Eq. (\ref{NMCfieldeqs}) and thus only valid asymptotically ({\it i.e.} for a late time de Sitter universe), this naturally occurs only when these two conditions are evaluated at $R = 4\La$ and the matter content of the Universe is only given by the vacuum energy $\La_0$:
\begin{equation}\label{decomp} \cl = -2 \ka \La_0 \to T_\mn = -2\ka \La_0 g_\mn \to T = -8\ka \La_0 ~~, \end{equation}
so that Eqs. (\ref{condition}) and (\ref{tracecondition}) read
\begin{equation}
f'_1(4\La) - 2f'_2(4\La) \La_0 = 1 ~~~~,~~~~
f_1(4\La) - 2 f_2(4\La) \La_0 = 2\La ~~.
\end{equation}

Thus, instead of resorting to cosmic averages or a fine-tuned integration constant, we conclude that the discrepancy between the values of $\La$ and $\La_0$ is imputed on the forms of the functions $f_1(R)$ and $f_2(R)$ defining the model. Notice that, in the case of GR, the above implies that $\La = \La_0$: there is no sequester of the vacuum energy contribution, so that the observed value of the CC should coincide with the later.

We now ascertain how this may be used to alleviate the CC problem. If we consider the effect of the curvature term to be similar to those of GR, $f_1(4\La) \approx R = 4\La$ and $f_1'(4\La) \approx 1$, we obtain

\begin{equation}
f'_2(4\La)\La_0 \ll 1~~~~,~~~~f_2(4\La)\La_0 \approx \La ~~.
\end{equation}
If we instead consider a feeble NMC, so that $f_2(4\La) \approx 1$ and $f_2'(4\La) \approx 0$, we obtain
\begin{equation}
f'_1(4\La) \approx 1 ~~~~,~~~~
f_1(4\La) = 2(\La_0 + \La) \approx 2\La_0~~.
\end{equation}

\section{Conclusions}

In this work we have established a mechanism through which one may equate the vacuum-energy contribution to the CC $\La_0$ with its observed value $\La$ via Eqs. (\ref{condition}) and (\ref{tracecondition}). This relation stems from the assumption that the cosmological dynamics have relaxed towards an asymptotic regime, which has been thoroughly characterized via the equivalent dynamical system in Ref. \cite{NMCcosmo2}.

We obtain conditions for the functions $f_1(R)$ and $f_2(R)$ that should be fulfilled in order to overcome the discrepancy between $\La_0$ and $\La$: by a criterious choice of these, no additive fine-tuning is required, as the orders of magnitude between the latter should appear as a dimensionless parameter of the theory (which may be expressed via the ratio between the characteristic mass scales typifying $f_1(R)$ and $ f_2(R)$ and $\La_0$). Given that the model under scrutiny (\ref{NMCaction}) is compatible with many inflationary models \cite{NMCinflation}, we conclude that the described mechanism might hint that a NMC might perhaps be an essential element of an effective model arising from a fundamental quantum gravity theory, and warrants further investigation due to its proficuous phenomenological consequences.

\end{document}